\documentclass [11pt]{article}
\usepackage{amsmath,amsthm,amsfonts,amscd,eucal,latexsym,amssymb}
\usepackage{empheq}
\usepackage{hyperref}
\usepackage{epsfig}  
\usepackage{color}
\newcommand{\normal}[1]{\mathop{:}\nolimits\!#1\!\mathop{:}\nolimits} 

\def\beq{\begin{eqnarray}}
\def\eeq{\end{eqnarray}}

\newcounter{proposition}[section]
\newcounter{theorem}[section]
\newcounter{lemma}[section]
\newcounter{definition}[section]
\newcounter{corollary}[section]
\def\theproposition{\thesection.\arabic{proposition}}
\def\thetheorem{\thesection.\arabic{theorem}}
\def\thelemma{\thesection.\arabic{lemma}}
\def\thedefinition{\thesection.\arabic{definition}}
\def\thecorollary{\thesection.\arabic{corollary}}

\newcommand{\se}[1]{\section{#1}}

\def\vsp{\vspace{0.2cm}}
\def\vspp{\vspace{0.1cm}}

\def\sse #1 {\vsp\ifhmode{\par}\fi\refstepcounter{subsection}
  \noindent {\bf\thesubsection}. {\it #1}.\quad
  \addcontentsline{toc}{subsection}{\protect\numberline{\thesubsection} #1}%
  }

\def\ssb #1 {\vsp\ifhmode{\par}\fi\refstepcounter{subsection}
  \noindent {\bf\thesubsection.} {\bf #1.}\quad
  \addcontentsline{toc}{subsection}{\protect\numberline{\thesubsection} #1}%
  }

\def\ssa #1 {\ifhmode{\par}\fi\refstepcounter{subsection}
  \noindent {\bf\thesubsection.} {\bf #1.}\quad
  \addcontentsline{toc}{subsection}{\protect\numberline{\thesubsection} #1}%
  }

\def\proposizione #1 {\vsp\ifhmode{\par}\fi\refstepcounter{proposition}
  \vsp\ifhmode{\par}\fi\noindent {\bf Proposition \theproposition}. \quad {\it #1}}
\def\teorema #1 {\vsp\ifhmode{\par}\fi\refstepcounter{theorem}
  \vsp\ifhmode{\par}\fi\noindent {\bf Theorem \thetheorem}. \quad {\it #1}}
\def\lemma #1 {\vsp\ifhmode{\par}\fi\refstepcounter{lemma}
  \vsp\ifhmode{\par}\fi\noindent {\bf Lemma \thelemma}. \quad {\it #1}}
\def\definizione #1 {\ifhmode{\par}\fi\refstepcounter{definition}
  \vsp\ifhmode{\par}\fi\noindent {\bf Definition \thedefinition}. \quad {\it #1}}
\def\corollario #1 {\vsp\ifhmode{\par}\fi\refstepcounter{corollary}
  \vsp\ifhmode{\par}\fi\noindent {\bf Corollary \thecorollary}. \quad {\it #1}}

\def\proof #1 {\vspp\ifhmode{\par}\fi\noindent {\it Proof.} {#1} $\Box$\vsp\par}
\def\remark {\vsp\ifhmode{\par}\fi\noindent\noindent {\bf Remark:} 
}

\begin{document}

\par 
\bigskip 
\par 
\rm 
 
\par 
\bigskip 
\LARGE 
\noindent 
{\bf Curvature fluctuations on asymptotically de Sitter spacetimes via the semiclassical Einstein's equations} 
\bigskip 
\par 
\rm 
\normalsize 

\large
\noindent {\bf Claudio Dappiaggi$^{1,2,a}$},
{\bf Alberto Melati$^{1,b}$},  \\
\par
\small
\noindent $^1$ 
Dipartimento di Fisica, Universit\`a degli Studi di Pavia,
Via Bassi, 6, I-27100 Pavia, Italy.\smallskip

\noindent$^2$  Istituto Nazionale di Fisica Nucleare - Sezione di Pavia,
Via Bassi, 6, I-27100 Pavia, Italy.

\bigskip

\noindent $^a$  claudio.dappiaggi@unipv.it,
$^b$  alberto.melati01@universitadipavia.it \\ 
 \normalsize

\par 
 
\rm\normalsize 
\noindent {\small Version of \today}

\rm\normalsize 
 
 
\par 
\bigskip 

\noindent 
\small 
{\bf Abstract}. 
It has been proposed recently to consider in the framework of cosmology an extension of the semiclassical Einstein's equations in which the Einstein tensor is considered as a random function. This paradigm yields a hierarchy of equations between the $n$-point functions of the quantum, normal ordered, stress energy-tensor and those associated to the stochastic Einstein tensor. Assuming that the matter content is a conformally coupled massive scalar field on de Sitter spacetime, this framework has been applied to compute the power spectrum of the quantum fluctuations and to show that it is almost scale-invariant. We test the robustness and the range of applicability of this proposal by applying it to a less idealized, but physically motivated, scenario, namely we consider Friedmann-Robertson-Walker spacetimes which behave only asymptotically in the past as a de Sitter spacetime. We show in particular that, under this new assumption and independently from any renormalization freedom, the power spectrum associated to scalar perturbations of the metric behaves consistently with an almost scale-invariant power spectrum.   
\normalsize
\bigskip

\se{Introduction}
Cosmology is rapidly arising as the new test-ground for the effectiveness of quantum field theory as the right framework to describe the matter constituents of the Universe. Most of the cosmological effects, which have been also confirmed experimentally, are almost unanimously accepted to be strictly correlated to a quantum description of matter. A prototype example of this assertion is the Sachs-Wolfe effect, according to which the redshift in the cosmic microwave radiation is caused by fluctuations in the gravitational field and correspondingly in the matter density \cite{SW}.  In the framework of inflation these are in turn seeded by the quantum fluctuations during the era of exponential expansion. Several explicit computations have been proposed in the literature, all aimed at reproducing a power spectrum, that is the Fourier transform of the two-point distribution of the underlying quantum state, which is close to, but not exactly scale invariant. 

Recently a novel, interesting, approach has been brought forth: Curvature fluctuations are calculated via the semiclassical Einstein's equations from the fluctuations of the stress-energy tensor of the underlying matter fields \cite{Pinamonti:2013zba}. More precisely the idea is to interpret the Einstein tensor not as a classical quantity, but rather as a stochastic field whose $n$-point momenta are taken as equal to the symmetrized $n$-point functions of the quantum stress energy tensor, evaluated on a Gaussian state of Hadamard form. In a few words the latter is a condition which ensure that, up to a suitable regularization of the physical observables, no divergence occurs. Although a computation of the power spectrum has been attempted in the framework of stochastic gravity -- see for example \cite{Verd}, the approach introduced in \cite{Pinamonti:2013zba} is radically different. Let us review the key aspects: One considers a cosmological/flat patch of de Sitter spacetime and a conformally coupled, massive scalar fields thereon. This is quantized following the standard prescription and the expectation value of its regularized stress energy tensor is computed with respect to the Bunch-Davies state, which is a maximally symmetric ground state. Subsequently the authors of \cite{Pinamonti:2013zba} consider scalar fluctuations of the metric and they compute via the semiclassical Einstein's equations the influence of quantum fluctuations. It is a noteworthy that the from of these perturbations is similar to the Bardeen potential in a suitable gauge fixing. The main result of \cite{Pinamonti:2013zba} is that the ensuing power spectrum tends for early times to a form consistent with that of Harrison-Zel'dovich. Most importantly the whole approach is not affected by any renormalization ambiguity and it relies only on the singular structure of the two-point function of the Bunch-Davies state, the so-called Hadamard form. 

Although the above approach is a toy model which cannot be directly connected to a specific model of inflation, the outcome of its analysis is rather striking, to the extent that the effectiveness of the result prompts a question on its robustness, which is the main point we want to address in this paper. More precisely, one could wonder how much the conclusions of \cite{Pinamonti:2013zba} are affected by the choice of having a global de Sitter spacetime. Notice that any perturbation from this maximally symmetric scenario would entail also the lack of a unique ground state such as the Bunch-Davies and, therefore, one might also wonder whether the overall construction reproduces similar results in a less idealized scenario. The goal of our investigation is to show that this is indeed the case. As a matter of fact we shall consider a large class of Friedmann-Robertson-Walker (FRW) spacetimes characterized by having a flat spatial section and a scale factor which tends only asymptotically in time to that of de Sitter. This assumption captures geometrically the idea that, during the inflationary era, the spacetime was in a {\em quasi} de-Sitter phase evolving at later times according to the description of the standard cosmological models. From a quantum field theory point of view the class of backgrounds, we consider, has been studied in great detail \cite{Dappiaggi:2008dk, Dappiaggi:2007mx} and it enjoys a distinguished property: Regardless of the exact form of the scalar factor it is possible to associate to a massive, conformally coupled scalar field living thereon, a distinguished ground state of Hadamard form, invariant under the action of all isometries. Furthermore each of these states tends asymptotically to the Bunch-Davies one, hence creating a clear bridge to the scenario considered in \cite{Pinamonti:2013zba}. 

In comparison to the exact de Sitter case, our investigation is considerably more complicated since the lack of a precise form for the scale factor in the metric entails that the power spectrum has to be built out of the retarded Green operator of an hyperbolic partial differential equations which we cannot solve exactly, but we can evaluate only via a Dyson series. Although, from a mathematical point of view, this is a legitimate operation, each order in the Dyson series contributes in turn to the power spectrum itself. We show that, while the leading order leads to a behaviour consistent with that of Harrison-Zel'dovich, the first order (and reiterating the same procedure, also all other orders) yields a sub-leading contribution which actually vanishes asymptotically in time. In other words we prove that the range of applicability of the method proposed in \cite{Pinamonti:2013zba} is wide and robust.

Our paper is organized as follows: In the next section we review the approach of \cite{Pinamonti:2013zba} and in particular how to use the semiclassical Einstein's equations to obtain the power spectrum. In section $3$, we review instead the analysis of \cite{Dappiaggi:2008dk, Dappiaggi:2007mx} showing, in the framework of algebraic quantum field theory, how one can associate a distinguished Hadamard Gaussian state to a massive, conformally coupled scalar field living on a FRW Universe which behaves asymptotically as the de Sitter spacetime. This section is more technical than the others, but it serves to the purpose of proving that all ingredients we need to compute the power spectrum do exist and are mathematically well-defined. In other words no assumption is made barring that on the form the metric. In section 4, we obtain our main result: the computation of the power spectrum as a perturbative series and the proof that the leading order behaves consistently with the Harrison-Zel'dovich spectrum while the higher orders are sub-dominant and do not contribute asymptotically in time. Eventually in section 5 we draw our conclusions.

\se{The semiclassical Einstein's equations in a stochastic framework}\label{sec2}
The goal of this section is to review the approach proposed in \cite{Pinamonti:2013zba}. The starting point is the following: Let us consider any globally hyperbolic four-dimensional, Lorentzian spacetime $(M,g)$ and let us consider thereon a conformally coupled, massive scalar field which obeys the equation 
\begin{equation}\label{KG}
K\phi=\left(\Box-\frac{R}{6}-m^2\right)\phi=0,
\end{equation}
where $\Box=g^{ab}\nabla_a\nabla_b$ is the standard wave operator, while $R$ is the scalar curvature associated to $g$. Furthermore we do not let the metric be fixed but we let it evolve via the semiclassical Einstein's equations 
\begin{equation}\label{semicl}
G_{ab}=\omega(\normal{T_{ab}}),
\end{equation}
where $\omega(\normal{T_{ab}})$ is the expectation value on the state $\omega$ of the Wick ordered stress-energy tensor associated to $\phi$. Implicitly we have set the Newton constant so that the factor $8\pi G$ on the right hand side of \eqref{semicl} is equal to $1$. Let us remark, that, on curved backgrounds, the construction of Wick polynomials is not obtained via normal ordering but via a point-splitting procedure followed by a subtraction of the relevant singular contributions \cite{Hollands:2001nf}. Such procedure, known as Hadamard regularization, relies on choosing $\omega$ within the special class of the so-called {\it Hadamard states}. In the next sections we will discuss the Hadamard condition more in detail, particularly in the case of cosmological spacetimes. For the moment we remark that $T_{ab}$ does not have the standard textbook form, but it includes an additional contribution proportional to $\frac{1}{3}g_{ab}\phi K\phi$. First introduced in \cite{Mo03}, it accounts for the fact that, otherwise, the divergence of $\omega(\normal{T_{ab}})$ would not vanish, hence yielding a contradiction since $\nabla^a G_{ab}=0$. The price to pay is the addition to $\normal{T}=\normal{g^{ab}T_{ab}}$ of a non trivial contribution also known as trace anomaly -- see again \cite{Mo03}. 

The semiclassical Einstein's equations are undoubtedly not valid in all regimes, but they are expected to provide a good understanding of the coupling between matter and gravity in an intermediate range where the first has to be described as a full-fledged quantum theory while the second retains essentially a classical description. While, on the one hand, such assertion can be verified only once a full theory of quantum gravity is understood and its behaviour in limiting regimes analyzed, on the other hand cosmology or rather the study of the evolution of the Universe in its early stages represents the natural playground for testing the effectiveness of \eqref{semicl}. An additional concern is raised by realizing that we are actually trying to equate a purely classical quantity, the left hand side, with an expectation value, hence a probabilistic quantity on the right hand side. This can only be meaningful if the quantum fluctuations of the Wick ordered stress-energy tensor are negligible. Yet, as duly noted in \cite{Pinamonti:2010is}, this is not the case as the variance of the unsmeared $\normal{T_{ab}}$ is always divergent. 

The possible way out from this quandary, proposed in \cite{Pinamonti:2013zba}, is to promote $G_{ab}$ to a random function, which thus allows also for random fluctuations on the left hand side. On his own this idea would not work, as the right hand side would still yield a divergent variance. Yet this problem can be bypassed assuming that \eqref{semicl} is just the first of a family of equations yielding a relation between the $n$-point functions of $\normal{T_{ab}}$ and that of Einstein tensor. In other words:
\begin{eqnarray}
\langle G_{ab}(x)\rangle&=&\omega(\normal{T_{ab}(x)}) \label{G_first}\\
\langle\delta G_{ab}(x)\delta G_{cd}(x^\prime)\rangle & = &\frac{1}{2}\omega\left(\normal{\delta T_{ab}(x)}\,\normal{\delta T_{cd}(x^\prime)}+\normal{\delta T_{cd}(x^\prime)}\,\normal{\delta T_{ab}(x)}\right)\notag\\
\langle(\delta G)^{\boxtimes n}(x_1,\ldots,x_n)\rangle & = &\mathfrak{S}\left[\omega(\normal{\delta T}^{\boxtimes n}(x_1,\ldots,x_n))\right], \quad n>1,\notag
\end{eqnarray}
where $\delta G_{ab}\doteq G_{ab}-\langle G_{ab}\rangle$ and $\normal{\delta T_{ab}}\doteq \normal{T_{ab}}-\omega(\normal{T_{ab}})$. Notice that $\boxtimes$ stands for the outer tensor product and that $\langle\cdot\rangle$ amounts to taking the average with respect to the underlying probability distribution. The symmetrization denoted by $\mathfrak{S}$ is a by-product of the invariance of the left hand side by permutations of the points $x_i$, $i=1,...,n$. As stressed in \cite{Pinamonti:2013zba}, the above hierarchy of equations allows in principle to compute, after smearing both sides, the moments of the smeared Einstein tensor. Nonetheless this is not tantamount to determining a probability distribution as the operation of extracting it from the associated momenta is not necessarily well-defined and it has been successfully implemented only in a few cases \cite{Fewster:2010mc, Fewster:2012ej}.

From a concrete point of view, we shall employ the above scheme of reasoning as follows: We consider a \emph{background spacetime} $(M,\overline{g})$ together with $(M,g)$, a \emph{perturbed spacetime}, which fluctuates around the former so that $\langle G_{ab}\rangle$ coincides with the Einstein tensor built out of $\overline{g}$ and $\omega(\normal{T_{ab}(x)})$ is built with respect to a quantum field theory living on the background spacetime. 

More specifically, as we will discuss thoroughly in Section \ref{sec4}, we are interested in a special scenario, namely cosmological backgrounds \`a la Friedmann-Robertson-Walker (FRW) and a matter content described by a massive, conformally coupled, scalar field. Suppose that we can associate to the latter a Gaussian Hadamard state, invariant under all background isometries, then, we can use the above set of equations to compute the perturbations of the background spacetime out of the right hand side. Since, as we will make explicit in the next section, an isotropic and homogeneous background has only one dynamical degree of freedom, we can actually simplify the above hierarchy of equations by considering the trace with respect to $\overline{g}$ in \eqref{G_first} and, for consistency, in the equations for all $n$-point function. If we introduce, therefore, the auxiliary variable $S(x)\doteq -\overline{g}^{ab}(x)G_{ab}(x)$ where $G_{ab}(x)$ is the stochastic Einstein tensor, we obtain \cite{Pinamonti:2013zba}
\begin{subequations}
\begin{flalign}
&\langle S(x)\rangle=m^2 [W]-2[V_1]-\alpha m^2, \label{traceanomaly}\\
&\langle S(x)S(x^\prime)\rangle-\langle S(x)\rangle\langle S(x^\prime)\rangle=m^4\left(\omega_2^2(x,x^\prime)+\omega_2^2(x^\prime, x)\right),\label{2-perturb}\\
& \langle(S-\langle S\rangle)^{\boxtimes n}(x_1,\ldots,x_n)\rangle=2^nm^{2n}\mathfrak{S}\left[\sum_{G}\prod_{i,j}\frac{\omega_2(x_i,x_j)^{\lambda_{ij}^G}}{\lambda_{ij}^G!}\right], \quad n>1,\label{n-perturb}
\end{flalign}
\end{subequations}
where $\omega_2(x,x^\prime)$ represents the two-point function of the Hadamard state $\omega$. The sum is over all directed graphs $G$ with $n$ vertexes and with two arrows at every vertex directed to other vertexes with higher labels. The quantity $\lambda_{ij}^G$ is equal to $0$, $1$ or $2$ depending on the number of arrows from the vertex $i$ to $j$-th one. Notice that the right hand side of \eqref{traceanomaly} is actually $\omega(\normal{T})$, which includes the contribution both from the trace anomaly and from the regularization freedom intrinsic in the Hadamard prescription. It has been already computed several times in the literature and, here, we refer in particular to \cite{Dappiaggi:2008mm}. Without entering too much into the details which are not of major relevance to our forthcoming analysis, we mention that $[\cdot]$ stands for the Synge bracket $[f](x)=f(x,x)$, $[W]$ is a state dependent contribution whereas $[V_1]$ is due to the trace anomaly \cite{Wald:1978pj}. Notice also that $\alpha$ is the only remaining renormalization constant and it is thus a priori undetermined. In the most generic scenario there are two additional constants which have been here implicitly set to $0$ either since we want have a well-defined initial value problem, canceling thus all terms proportional to $\Box R$, or since we can reabsorb all terms proportional to $m^2 R$ in the definition of the Newton constant. 

\se{Scalar quantum field theory on asymptotically de Sitter spacetimes}\label{sec_3}
In order to compute the perturbations of the geometry out of \eqref{traceanomaly}, \eqref{2-perturb} and \eqref{n-perturb} the first step is to make sure that to the matter content, we are interested in, we can associate a Gaussian Hadamard state, which is also invariant under all background isometries. In \cite{Pinamonti:2013zba} only a massive, conformally coupled scalar field on de Sitter spacetime has been considered and, in this case, one can use the so-called Bunch-Davies state which enjoys all the desired properties \cite{Allen:1985ux, BD, Schomblond:1976xc}. 

Hence goal of this section will be to introduce a class of spacetimes which generalizes the cosmological de Sitter spacetime showing at the same time how to quantize thereon a massive scalar field, conformally coupled to scalar curvature and how to assign to it a Gaussian Hadamard state. The material of this section is based mainly on \cite{Dappiaggi:2008dk, Dappiaggi:2007mx}. We shall try to keep the details to the bare minimum needed to make this paper self-consistent and, hence, also the presentation of the quantization procedure of a field theory on curved backgrounds will be only outlined and adapted to the case at hand. For more details we refer the reader to recent reviews \cite{Benini:2013fia, Hollands:2014eia}. 

Let us consider as background an homogeneous and isotropic four dimensional smooth manifold $M$, together with a Lorentzian metric $g$. In a local coordinate patch $(\tau,r,\theta,\varphi)$ the line element reads $ds^2=a^2(\tau)\left(-d\tau^2+dr^2+r^2d\mathbb{S}^2(\theta,\varphi)\right)$, where $r\in (0,\infty)$ while $d\mathbb{S}^2(\theta,\varphi)$ is the standard metric on the unit $2$-sphere. The variable $\tau$ takes values in an interval $I\subseteq\mathbb{R}$ while $a(\tau)$ is a smooth and strictly positive function, also dubbed scale factor. In other words we are considering a Friedmann-Robertson-Walker (FRW) spacetime with flat spatial section whose metric is written in the conformal time. Following \cite{Dappiaggi:2008dk, Dappiaggi:2007mx}, we assume that $\tau\in (-\infty,0)$ and that 
\beq\label{conf_fac}
a(\tau)=-\frac{1}{H\tau}+O\left(\frac{1}{\tau^2}\right);\qquad \frac{da(\tau)}{d\tau}=\frac{1}{H\tau^2}+O\left(\frac{1}{\tau^3}\right)\qquad \frac{d^2a(\tau)}{d\tau^2}=-\frac{2}{H\tau^3}+O\left(\frac{1}{\tau^4}\right),
\eeq
where $H>0$ is a constant and where all expressions are meant as asymptotic expansions as $\tau\to -\infty$. Notice that, from a mere technical point of view, we could work without key differences with the conformal time defined in the interval $(0,+\infty)$ provided that the sign in front of $(H\tau)^{-1}$ is positive. Our choice is dictated by a physical reasoning: First of all notice that, if we introduce the standard cosmological time $t$ via the defining relation $dt=a(\tau)d\tau$ and if we assume that $a(\tau)=-\frac{1}{H\tau}$, the metric becomes $ds^2=-dt^2+e^{2Ht}\left(dr^2+r^2d\mathbb{S}^2(\theta,\varphi)\right)$. In other words $H$ plays the role of the Hubble constant and this is the standard de Sitter Universe. Hence the asymptotic constraint \eqref{conf_fac} entails that we are considering FRW spacetimes which tend to the de Sitter one at early times $t$, hence as $\tau\to -\infty$.

We stress that the choice of \eqref{conf_fac} is motivated both from a physical and from a mathematical point of view. As a matter of fact, it is common wisdom in modern cosmological theories, accepting the inflationary paradigm, that the phase of exponential expansion which the Universe has undergone in its early stages is best described by an almost de Sitter metric and our hypothesis captures at a geometric level such statement. 

From a mathematical point of view the class of cosmological backgrounds, that we are considering, enjoys several notable properties. To start with all FRW spacetimes are globally hyperbolic and, as a consequence, all constant time $\tau$ submanifolds are Cauchy hypersurfaces. Thereon it is possible to assign initial data for any partial differential equation of hyperbolic type such as the massive Klein-Gordon field conformally coupled to scalar curvature, which we consider in this paper. A more distinguished feature which descends from \eqref{conf_fac} and which has been thoroughly investigated in \cite{Dappiaggi:2008dk} is that all spacetimes under analysis possess a boundary, which is a null three-dimensional submanifold which we call $\Im^-$. It plays the role of the cosmological horizon as defined in \cite{Rindler} and it can be characterized introducing the coordinates $U=\tan^{-1}(\tau-r)$ and $V=\tan^{-1}(\tau+r)$. By direct substitution one can realize that $(M,g)$ can be read as an open submanifold of a larger spacetime $(\widetilde M,\widetilde g)$ which is still globally hyperbolic and $\widetilde g|_M= g$. In this picture $\Im^-=\partial J^+(M;\widetilde M)$ where $J^+$ denotes the causal future. Although such boundary enjoys several additional properties which mimic that of null infinity on an asymptotically flat spacetimes, it is its very existence which makes the class of backgrounds under investigation rather distinguished. We shall not give thus further geometric details and we shall discuss instead how to quantize a scalar field theory on $(M,g)$ and why $\Im^-$ plays a key role.

Let us consider $\phi:M\to\mathbb{R}$ a real scalar field which obeys \eqref{KG}. Since $K$ is a normally hyperbolic operator, we can introduce 
$$\mathcal{S}(M)=\left\{\phi\in C^\infty(M)\;|\;\exists f\in C^\infty_0(M)\;\textrm{such that}\; \phi=E(f)\right\},$$
where $E=E^+-E^-$ is the causal propagator, hence the difference between the advanced ($E^+$) and the retarded ($E^-$) fundamental solutions/Green operators. These are maps $E^\pm: C^\infty_0(M)\to C^\infty(M)$ such that $K\circ E^\pm= E^\pm \circ K =id$, where $id$ is the identity, and such that $supp(E^\pm(f))\subseteq J^\pm(supp(f))$ for all $f\in C^\infty_0(M)$. Here $J^\pm$ is the causal future (+) / past (-). $\mathcal{S}(M)$ represents the set of all solutions of \eqref{KG} with a compactly supported intersection with any Cauchy surface, that is, in our case, the constant $\tau$ hypersurfaces. Although this is a reduced class of all possible smooth solutions of \eqref{KG} it suffices to construct a quantum algebra of observables for the Klein-Gordon field. As a starting point we endow $\mathcal{S}(M)$ with a weakly non-degenerate symplectic form $\sigma_M$, namely:
$$\sigma_M(\phi,\phi^\prime)=E(f,f^\prime)=\int\limits_M d\mu(g) E(f)(x)f^\prime(x),$$
where $d\mu(g)$ is the metric induced volume form while $f,f^\prime\in C^\infty_0(M)$ are such that $\phi=E(f)$ and $\phi^\prime=E(f^\prime)$. By following a standard construction, we associate an {\it algebra of fields} $\mathcal{F}(M)$ to the Klein-Gordon field fulfilling \eqref{KG} as follows 
\begin{enumerate}
\item Consider $C^\infty_0(M;\mathbb{C})=C^\infty_0(M)\otimes\mathbb{C}$ and $\mathcal{T}(M)\doteq\bigoplus_{n=0}^\infty C^\infty_0(M;\mathbb{C})^{\otimes n}$ where we set the $n=0$ term to be $\mathbb{C}$. The tensor product endows $\mathcal{T}(M)$ of an algebra structure which can be supplemented with the $*$-operation given by complex conjugation,
\item Construct the $*$-ideal $\mathcal{I}(M)$ generated by elements of the form $Pf$ and $f\otimes f^\prime - f^\prime\otimes f - i E(f,f^\prime)\mathbb{I}$, where $\mathbb{I}$ is the unit element in $\mathcal{T}(M)$.
\item Define the algebra of fields or, in other words, the set of all possible local observables as 
$$\mathcal{F}(M)\doteq\frac{\mathcal{T}(M)}{\mathcal{I}(M)}.$$
\end{enumerate}
In order to have a full-fledged quantization scheme, it is necessary to assign to the algebra of fields a quantum state that is a positive functional $\omega:\mathcal{F}(M)\to\mathbb{C}$. Via the GNS theorem it is possible to recover from $(\mathcal{F}(M),\omega)$ the standard probabilistic interpretation of a quantum theory -- see for example \cite{Benini:2013fia}. In the class of all possible states we shall restrict our attention first of all to those which are quasi-free/Gaussian, that is all $n$-point functions can be constructed out of the $2$-point one as follows: Let $f_1,...,f_n\in C^\infty_0(M)$, then, for $n$ even, 
\begin{equation}\label{n-pt}
\omega_n(f_1,...,f_n)=\sum\limits_{\pi_n\in S^\prime_n}\prod\limits_{i=1}^{n/2}\omega_2(f_{\pi_n(2i-1)},f_{\pi_n}),
\end{equation}
where $S^\prime_n$ denotes the set of ordered permutations of $n$-elements, while $\omega_2$ is the two-point function. On the contrary, if $n$ is odd, then $\omega_n$ is set to $0$. In other words specifying a quasi-free state $\omega$ on $\mathcal{F}(M)$ is tantamount to assigning $\omega_2$ fulfilling a positivity condition. 

Between the plethora of all possible Gaussian quantum states, not all of them are considered physically acceptable, since, in many cases, unwanted pathologies do occur. The prime example is the blow-up of the quantum fluctuations of certain observables such as the smeared components of the stress-energy tensor. In order to avoid these issues, it has been unveiled that a good criterion is to require that $\omega$ fulfills the so-called microlocal spectrum condition, which turns out to be equivalent to imposing a constraint on the singular structure of $\omega_2$. These special states are also dubbed {\it Hadamard states}. Although we do not dwell into analyzing their properties in detail, we refer a reader to \cite{Benini:2013fia} for further details. It is noteworthy that, for a given free quantum field theory and for a given background, there exist infinite possible Hadamard states, although a distinguished subclass is represented by those which are invariant under the action of the isometries of the underlying spacetime. For a massive Klein-Gordon field with an arbitrary coupling to curvature, the canonical example is the Poincar\'e vacuum in Minkowski spacetime or the so-called Bunch-Davies state on de Sitter spacetime \cite{Allen:1985ux, BD, Schomblond:1976xc}, which has been at the heart of the analysis in \cite{Pinamonti:2013zba}.

Since we are interested in working on FRW spacetimes $(M,g)$ with a conformal factor constrained by \eqref{conf_fac}, we outline how it is possible thereon to construct a Hadamard state, invariant under isometries. Notice that failing to build such a state would entail that, per consistency, there could not exist a solution to the semiclassical Einstein's equations. For the rest of the section we summarize the procedure described in \cite{Dappiaggi:2008dk, Dappiaggi:2007mx}. The key idea is to exploit that $(M,g)$ is actually an open subset of a larger globally hyperbolic spacetime $(\widetilde M,\widetilde g)$, which contains $\Im^-$, the null three-dimensional manifold introduced above. This feature allows to employ a so-called bulk-to-boundary correspondence which consists of mapping $\mathcal{F}(M)$ into an auxiliary algebra intrinsically constructed on the cosmological horizon. The latter, which we call $\mathcal{F}(\Im^-)$ is built as follows: Consider $\Im^-$, which is a null differentiable manifold topologically $\mathbb{R}\times\mathbb{S}^2$ and, in analogy to the construction of null infinity on asymptotically flat spacetimes, endow $\Im^-$ with the coordinates $(u,\theta,\varphi)$ so that the metric $\widetilde g$ reads on $\Im^-$ as follows:
$$\widetilde{ds}^2|_{\Im^-}=H^{-2}(-2duda+d\mathbb{S}^2(\theta,\varphi)).$$
Here $(\theta,\varphi)$ are the standard coordinates on the unit $2$-sphere while $u$ is the affine parameter of the vector field $n_\mu=\nabla_\mu a(\tau)$, which turns out to be extensible to a null complete geodesic for $\widetilde g$ on $\Im^-$. Let us introduce the following space of functions:
$$\mathcal{S}(\Im^-)=\left\{\psi\in C^\infty(\Im^-)\;|\;||\psi||_\infty,||k\widehat{\psi}||_\infty<\infty\;\partial_u\psi\in L^1(\mathbb{R}\times\mathbb{S}^2,dld\mathbb{S}^2),\;\widehat{\psi}\in L^1(\mathbb{R}\times\mathbb{S}^2;dkd\mathbb{S}^2)\right\}$$
where $\widehat{\psi}(k,\theta,\varphi)\doteq\int\limits_{\mathbb{R}}\frac{dk}{\sqrt{2\pi}} \psi(u,\theta,\varphi)e^{iku}$ and $||\cdot||_\infty$ stands for the $L^\infty$-norm. Notice that $\mathcal{S}(\Im^-)$ is actually a symplectic space if endowed with the following weakly non degenerate antisymmetric bilinear form:
$$\sigma_\Im(\psi,\psi^\prime)=\int\limits_{\mathbb{R}\times\mathbb{S}^2}dud\mathbb{S}^2\left(\psi\frac{\partial\psi^\prime}{\partial u}-\psi^\prime\frac{\partial\psi}{\partial u}\right).$$
Following the same procedure as for $(\mathcal{S}(M),\sigma_M)$ it is possible to associate to $(\mathcal{S}(\Im^-),\sigma_\Im)$ a $*$-algebra as follows:
\begin{itemize}
\item Consider $\mathcal{S}(\Im^-)_\mathbb{C}\doteq\mathcal{S}(\Im^-)\otimes\mathbb{C}$ and let $\mathcal{T}(\Im^-)=\bigoplus_{n=0}^\infty \mathcal{S}(\Im^-)_\mathbb{C}^{\otimes n}$ where we define the term $n=0$ as $\mathbb{C}$. This is a $*$-algebra if endowed with the complex conjugation as $*$-operation.
\item We quotient a $*$-ideal $\mathcal{I}(\Im^-)$ generated by elements of the form $\psi\otimes\psi^\prime-\psi^\prime\otimes\psi-i\sigma_\Im(\psi,\psi^\prime)I$ where $I$ is the identity element in $\mathcal{T}(\Im^-)$. Accordingly we define 
$$\mathcal{F}(\Im^-)=\frac{\mathcal{T}(\Im^-)}{\mathcal{I}(\Im^-)}.$$
\end{itemize}
Notice that the algebra on the boundary plays only an auxiliary role and it has no intrinsic physical meaning. As a matter of fact it does not contain any information on the dynamics in $(M,g)$ as one can infer by noticing that $\mathcal{I}(\Im^-)$ is built only out of the canonical commutation relations written out of the symplectic form, contrary to what happens for $\mathcal{I}(M)$.
In order to relate $\mathcal{F}(M)$ to $\mathcal{F}(\Im^-)$ one has to make use of the following fact: $(M,g)$ can be read as an open subset of a larger globally hyperbolic spacetime $(\widetilde M,\widetilde g)$. For this reason, consider any $f\in C^\infty_0(M)$ and construct $E_{\widetilde M}(f)$ where $E_{\widetilde M}$ is the causal propagator of $\Box-\frac{R}{6}-m^2$ on $\widetilde{M}$. Notice that, since $\widetilde g|_M=g$, the uniqueness of the solutions of the Klein-Gordon equation entails that $E_{\widetilde M}(f)=E(f)$ on $(M,g)$. In \cite{Dappiaggi:2008dk, Dappiaggi:2007mx} it has been proven the following:

\teorema{Let us consider $\Gamma:C^\infty_0(M)\to C^\infty(\Im^+)$ such that $\Gamma(f)=\left.E_{\widetilde M}(f)\right|_{\Im^-}$ where $|_{\Im^-}$ stands for the restriction to the cosmological horizon. Then  
\begin{itemize}
\item $\Gamma$ defines an injective map $\widetilde\Gamma: \mathcal{S}(M)\to\mathcal{S}(\Im^-)$ such that $\widetilde\Gamma(\phi)=\Gamma(f)$ where $f\in C^\infty_0(M)$ is such that $\phi=E(f)$. Furthermore $\widetilde\Gamma(\phi)=0$ if and only if $\phi=0$,
\item $\widetilde\Gamma$ preserves the symplectic forms that is, for all $\phi,\phi^\prime\in\mathcal{S}(M)$, 
$$\sigma_M(\phi,\phi^\prime)=\sigma_\Im(\widetilde\Gamma(\phi),\widetilde\Gamma(\phi^\prime)).$$
\end{itemize}}

As a byproduct, this theorem entails the existence of an injective homomorphism $\iota:\mathcal{F}(M)\to\mathcal{F}(\Im^+)$ which can be defined simply by stating its action on each $f\in C^\infty_0(M)$ that is $\iota(f)\doteq\widetilde\Gamma(E(f))$. In turn this map entails that for every boundary state, namely a positive functional $\omega_\Im:\mathcal{F}(\Im^-)\to\mathbb{C}$, one can associate via pull-bacl a counterpart $\omega_M:\mathcal{F}(M)\to\mathbb{C}$ such that, for every element $a\in\mathcal{F}(M)$
$$\omega_M(a)\doteq(\iota^*\omega_\Im)(a)=\omega_\Im(\iota(a)).$$
The advantage of this abstract procedure is that, in the case at hand, it is possible to exploit the peculiar geometric structure of the cosmological horizon to identify a distinguished Gaussian state for the auxiliary boundary theory. As explained in \eqref{n-pt} such state is defined in terms of its two-point function which reads:
$$\omega_{2,\Im}(\psi,\psi^\prime)=\lim\limits_{\epsilon\to 0}-\frac{1}{\pi}\int\limits_{\mathbb{R}^2\times\mathbb{S}^2}dudu^\prime d\mathbb{S}^2(\theta,\varphi)\;\frac{\psi(u,\theta,\varphi)\psi^\prime(u^\prime,\theta,\varphi)}{(u-u^\prime-i\epsilon)^2},$$
for all $\psi,\psi^\prime\in\mathcal{S}(\Im^-)$. An investigation of this state, completed mainly in \cite{Dappiaggi:2008dk} unveiled the following facts which are relevant to our analysis:
\begin{itemize}
\item The two-point function $\omega_{2,\Im}$ induced via pull-back a bulk counterpart reads as follows: For all $f,f^\prime\in C^\infty_0(M)$,
\begin{gather}\label{2-pt}
\omega_{2,M}(f,g)=\lim\limits_{\epsilon\to 0}-\frac{1}{\pi}\int\limits_{\mathbb{R}^2\times\mathbb{S}^2}dudu^\prime d\mathbb{S}^2(\theta,\varphi)\;\frac{\widetilde\Gamma(E(f))(u,\theta,\varphi)\widetilde\Gamma(E(f^\prime))(u^\prime,\theta,\varphi)}{(u-u^\prime-i\epsilon)^2},\\
\omega_{2,M}(f,g)=\int\limits_{\mathbb{R}\times\mathbb{S}^2}dkd\mathbb{S}^2(\theta,\varphi)\;2k\Theta(k)\overline{\widehat{\widetilde\Gamma(E(f))}(k,\theta,\varphi)}\widehat{\widetilde\Gamma(E(f^\prime))}(k,\theta,\varphi),\notag 
\end{gather}
where, in the second expression, we have performed the Fourier transform along the $u$-direction of both $\psi$ and $\psi^\prime$. Here $\Theta(k)$ is the Heaviside function.
\item The state $\omega_M$ for $\mathcal{F}(M)$ constructed out of \eqref{2-pt} is Gaussian and invariant under the action of all isometries of $(M,g)$. Furthermore $\omega_M$ is of Hadamard form and it coincides with the Bunch-Davies vacuum when the scale factor $a(\tau)=-\frac{1}{H\tau}$. 
\end{itemize}

To conclude the section, we stress once more that the Hadamard states constructed out of \eqref{2-pt} are of relevance not for their specific form, but rather for their very existence. In our analysis we will rather use the explicit form of the singular structure of the two-point function of a Hadamard state in a geodesic neighbourhood of a point of $(M,g)$. Yet, although one can in principle use deformation arguments -- see \cite{Fulling:1981cf} -- to prove the existence of states of Hadamard form in every globally hyperbolic spacetime, it is not obvious that these behave well under action of the isometries. The method depicted above dispels such doubt and it shows, moreover, that the outcome is a natural generalization of the Bunch-Davies state on de Sitter spacetime. For the sake of completeness we stress that a similar result could have been obtained by looking at the states of low-energy as constructed in \cite{Olbermann:2007gn}. Their relation to the Bunch-Davies state and more generally to those constructed in \cite{Dappiaggi:2008dk, Dappiaggi:2007mx} has been unveiled recently in \cite{Degner}.

\se{Fluctuations around an asymptotically de Sitter spacetime}\label{sec4}
On account of our discussion in Section \ref{sec2} and of the hypothesis on the form of the scale factor in \eqref{conf_fac} for the FRW metric $\overline{g}$ with flat spatial sections, we can consider scalar fluctuations of $\overline{g}$ so to obtain a perturbed metric $g$ whose line element reads:
\begin{equation}\label{fluct}
ds^2=a^2(\tau)\left[-\left(1+2\Psi\right)d\tau^2+\left(1-2\Psi\right)\left(dr^2+r^2d\mathbb{S}^2(\theta,\varphi)\right)\right],
\end{equation}
where $\Psi$ is a smooth scalar function. The form \eqref{fluct} appears very similar to the one appearing in the single-field models of inflation in absence of anisotropic stress and in the longitudinal gauge. In this case the two Bardeen potentials coincide and \eqref{fluct} is used \cite{Bardeen:1980kt}. Following the road paved in \cite{Pinamonti:2013zba}, we are interested, as explained in Section \ref{sec2}, in the perturbations of the Einstein tensor induced by $\Psi$. On account of the hypothesis of homogeneity and isotropy, we can avoid working with the whole $G_{\mu\nu}$ and we can instead consider the auxiliary quantities $S\doteq- \overline{g}^{\mu\nu} G_{\mu\nu}$, $G_{\mu\nu}$ being the Einstein tensor built out of $g$. If we call $\langle S\rangle$ the trace of the unperturbed Einstein tensor we obtain that 
\begin{equation}\label{eq_dyn}
S-\langle S\rangle=Q\Psi=-\frac{6}{a^4}\left(\frac{\partial^2}{\partial\tau^2}-\frac{1}{3}\nabla^2 + V(\tau)\right)a^2\Psi,
\end{equation}
where 
\begin{equation}\label{potential}
V(\tau)=2\left(\frac{\ddot{a}}{a}-2\frac{\dot{a}^2}{a^2}\right).
\end{equation}
Contrary to the de Sitter case, \eqref{eq_dyn} is not only the rescaling of a wave operator with an effective square velocity of $\frac{1}{3}$, but it includes also a non-trivial time-dependent potential such that, on account of \eqref{conf_fac}, $V(\tau)\mapsto 0$ as $\tau\to-\infty$. It is also worth remarking that, if we introduce the Hubble function in the conformal time $\mathcal{H}(\tau)=\frac{\dot{a}(\tau)}{a^2(\tau)}$, the potential reads also $V(\tau)=2a(\tau)\dot{\mathcal{H}}(\tau)$. 
Following the general strategy outlined in Section \ref{sec2}, we can translate equation \eqref{2-perturb} via \eqref{eq_dyn} to the 2-point correlation function for $\Psi$:
\begin{equation}\label{2-pt-pert}
\langle\Psi(x_1)\Psi(x_2)\rangle=m^4\int\limits_{M\times M}d^4y_1d^4y_2\;\Delta_R(x_1,y_1)\Delta_R(x_2,y_2)\left(\omega^2_2(y_1,y_2)+\omega^2_2(y_2,y_1)\right),
\end{equation}
where $\Delta_R$ stands for the retarded Green function of \eqref{eq_dyn} and where we have used implicitly that $\langle\Psi\rangle=0$. Notice that \eqref{2-pt-pert} entails that the correlations between the scalar fluctuations of the metric at two separate spacetime points (the left hand side) are ruled by the quantum fluctuations of the matter fields, encoded in $\omega_2$ in the right hand side.
 Since the spacetime metric is manifestly invariant under rotations and translations on all hypersurfaces at constant $\tau$, \eqref{2-pt-pert} depends actually only on $\vec{x}_1-\vec{x}_2$, the difference between the space components of $x_1$ and $x_2$. Hence the power spectrum at time $\tau$ can be defined via a Fourier transform as:
$$\langle\Psi(\tau,\vec{x}_1)\Psi(\tau,\vec{x}_2)\rangle=\frac{1}{(2\pi)^3}\int\limits_{\mathbb{R}^3}d^3k\;\mathcal{P}(\tau,\vec{k})e^{i\vec{k}\cdot (\vec{x}_1-\vec{x}_2)},$$
where the dot stands for the standard Euclidean product on $\mathbb{R}^3$. Upon Fourier transform, it descends 
\begin{equation}\label{Powerspectrum}
\mathcal{P}(\tau,\vec{k})=2m^4\int\limits_{-\infty}^\tau\int\limits_{-\infty}^\tau d\tau_1d\tau_2\;\widehat\Delta_R(\tau,\tau_1,\vec{k})\widehat\Delta_R(\tau,\tau_2,\vec{k})\widehat{\omega}^2_2(\tau_1,\tau_2,\vec{k}),
\end{equation}
where the symmetrization of the state in \eqref{2-pt-pert} has been taken care of by having the same limit of integration with respect to $\tau_1$ and $\tau_2$. In order to evaluate the last expression we need to provide, if not explicitly, at least in a controlled way, both $\omega_2$ and $\Delta_R$. While on the de Sitter spacetime, this can be readily achieved, we are facing a somehow more complicated scenario on account of the non-trivial form of the scale factor $a(\tau)$.  Hence we divide our analysis in two parts:

\vskip .2cm

\noindent {\it Evaluation of $\Delta_R$:} We are interested in analyzing the behaviour of the retarded fundamental solution of \eqref{eq_dyn}. As a starting point we consider the auxiliary equation $P\chi_k(\tau)=0$ where $P=\frac{\partial^2}{\partial\tau^2}+\frac{k^2}{3}+V(\tau)$ and $k$ here stands for the modulus of $\vec{k}$. The idea is to write down a solution $\chi_k(\tau)$ as a Dyson series in which $V(\tau)$ acts as a perturbation. In other words,
\begin{gather}
\chi_k(\tau)=\chi^0_k(\tau)+\sum\limits_{n=1}^\infty(-1)^n\int\limits_{-\infty}^\tau dt_1\int\limits_{-\infty}^{t_1} dt_2...\int\limits_{-\infty}^{t_{n-1}} dt_n\;\widehat{\Delta}_R^0(\tau,t_1,k)\widehat{\Delta}_R^0(t_1,t_2,k)...\widehat{\Delta}_R^0(t_{n-1},t_n,k)\notag\\
V(t_1)...V(t_n)\chi^0_k(t_n),\label{pert_series}
\end{gather} 
where $\chi^0_k$ is a solution of the unperturbed equation, {\it i.e.},
\begin{equation}\label{unp_Green}
\chi^0_k(\tau)=\int\limits_{\mathbb{R}}d\tau_1\;\widehat\Delta^0_R(\tau,\tau_1,k)f(\tau_1)\quad\widehat\Delta^0_R(\tau,\tau_1,k)=\frac{\sqrt{3}}{k}\sin\left(\frac{k(\tau-\tau_1)}{\sqrt{3}}\right),
\end{equation}
where $f(\tau_1)\in C^\infty_0(\mathbb{R})$ is an arbitrary initial datum. Notice that, on account of the fall-off behaviour of the scale factor and thus of $V(\tau)$, one can prove that there exists an interval $J=(-\infty, a)$ for which \eqref{pert_series} is absolutely convergent. This follows using, mutatis mutandis, the same arguments used in \cite[Appendix A]{Dappiaggi:2008dk}. We will not dwell into the details since it would lead us far from the main scope of the paper.

From \eqref{pert_series} we can, furthermore, extract the following series for the spatially Fourier-transformed retarded Green operator:
\begin{gather}
\widehat\Delta_R^V(\tau,\tau^\prime)=\widehat\Delta_R^0(\tau,\tau^\prime)+\sum\limits_{n=1}^\infty (-1)^n\int\limits_{-\infty}^\tau dt_1\int\limits_{-\infty}^{t_1}dt_2...\int\limits_{-\infty}^{t_{n-1}}dt_n\; \widehat\Delta_R^0(\tau,t_1)\widehat\Delta_R^0(t_1,t_2)\; ...\notag\\
\widehat\Delta_R^0(t_{n-1},t_n) V(t_1)\;...\;V(t_n)\widehat\Delta_R^0(t_n,\tau^\prime)\label{pert_series_Green}
\end{gather}
From this last expression we can construct for the operator $Q$ on the right hand side of \eqref{eq_dyn} the retarded Green operator, Fourier transformed along the spatial directions, as
\begin{equation}\label{ret_Green}
\widehat\Delta_R(\tau,\tau^\prime,k)=-\frac{a^4(\tau^\prime)}{6a^2(\tau)}\widehat\Delta_R^V(\tau,\tau^\prime,k),
\end{equation} 
where we have restored the explicit dependence on $k$, the modulus of $\vec{k}$. To prove the correctness of our formula, notice that, on the one hand, the structural properties of $\widehat\Delta_R$ as a map are inherited from those of $\widehat\Delta_R^V$, while, on the other hand, it is a right inverse of $\widehat Q$ (the spatial Fourier transform of $Q$) since
\begin{eqnarray*}
\widehat Q\widehat\Delta_R=\frac{6}{a^4(\tau)}Pa^2(\tau)\frac{a^4(\tau^\prime)}{6a^2(\tau)}\widehat\Delta_R^V=\frac{a^4(\tau^\prime)}{a^4(\tau)}P\widehat{\Delta}_R^V=\frac{a^4(\tau^\prime)}{a^4(\tau)}\delta(\tau-\tau^\prime)=\delta(\tau-\tau^\prime).
\end{eqnarray*}
An almost identical computation shows that $\widehat\Delta_R$ is also a left inverse of $\widehat Q$ and, hence, it enjoys all properties of a retarded Green operator. For future reference it is useful to express the retarded Green operator for $\widehat Q$ truncated at first order:
\begin{equation}\label{Green_1st}
\widehat\Delta_R^{(1)}(\tau,\tau^\prime,k)=-\frac{a^4(\tau_1)}{6a^2(\tau)}\widehat\Delta^0_R(\tau,\tau^\prime,k)+\frac{a^4(\tau_1)}{6a^2(\tau)}\int\limits_{-\infty}^\tau dt\;\widehat\Delta^0_R(\tau,t,k)V(t)\widehat\Delta^0_R(t,\tau^\prime,k).
\end{equation}

\vskip .2cm

\noindent {\it Evaluation of $\widehat\omega_2$:} The second ingredient that we need to control in order to evaluate \eqref{Powerspectrum} is the two-point function $\widehat\omega_2(\tau_1,\tau_2,\vec{k})$ for the underlying matter field, that is a massive, conformally coupled, Klein-Gordon field. In the previous section we have discussed the existence of Gaussian states which are of so-called Hadamard form and invariant under the action of all spacetime isometries. Their existence guarantees us that there is a natural candidate to work with, provided that the scale factor behaves asymptotically as in \eqref{conf_fac}. Yet, contrary to the Bunch-Davies state employed in \cite{Pinamonti:2013zba}, we cannot write down an explicit expression for the two-point function. It turns out that this is not a major hurdle since in order to compute the influence of the quantum matter on $\Psi$ via \eqref{2-pt-pert} or, more properly, the associate power spectrum \eqref{Powerspectrum}, it suffices to control the leading singularity of $\widehat\omega_2$. To this avail the construction of Section \ref{sec_3} has to be supplemented with the following property enjoyed by all Hadamard states: For any pair of point $x,x^\prime$ lying in a geodesically complete neighbourhood, the integral kernel of the two-point function reads \cite{Kay:1988mu}:
$$\omega_2(x,x^\prime)=\lim\limits_{\epsilon\to 0}\frac{1}{8\pi^2}\frac{U(x,x^\prime)}{\sigma_\epsilon(x,x^\prime)}+V(x,x^\prime)\ln\frac{\sigma_\epsilon(x,x^\prime)}{\lambda^2}+W(x,x^\prime),$$
where $\sigma_\epsilon\doteq \sigma+2i\epsilon(T(x)-T(x^\prime))+\epsilon^2$ is constructed out of $\sigma$, the halved squared geodesic distance between $x$ and $x^\prime$, $T$, a future increasing global time function and $\epsilon$ a regularization parameter. Furthermore $\lambda$ is an arbitrary reference scale length, whereas $U, V, W$ are smooth functions. While the first two are dependent only on local geometric quantities and on the parameters defining the underlying dynamics such as the mass and the coupling to scalar curvature, the latter can be chosen freely, up to the condition that $\omega_2(x,x^\prime)$ is a weak solution of \eqref{KG} both in $x$ and $x^\prime$. We have thus introduced the so-called local Hadamard form, which unveils in particular that the singular behaviour of the two-point function is controlled by the Hadamard parametrix $H(x,x^\prime)=\frac{1}{2\pi^2}\frac{U(x,x^\prime)}{\sigma_\epsilon(x,x^\prime)}+V(x,x^\prime)\ln\frac{\sigma_\epsilon(x,x^\prime)}{\lambda^2}$ which can be constructed fully out of the metric and of the equation ruling the dynamics. Notice, moreover, both that all singularities are localized at those points $x$ and $x^\prime$ which are either coincident or connected by a lightlike geodesic, for which $\sigma$ vanishes and that the logarithmic term has a sub-leading divergence compared to the other one. In other words, since we are interested in the leading singularity to evaluate the power spectrum, we can restrict our attention to the first term in $H(x,y)$. We can now make use of an additional important piece of our puzzle; the metric $\overline{g}$ is conformally related to that of Minkowski spacetime. Although the geodesic distance has a non trivial behaviour under a conformal transformation and thus one might not expect to be able to relate easily the local Hadamard forms in two conformally associated spacetimes, the following result is very handy and it follows simply adapting at the case at hand the calculations of \cite[App. A]{Pinamonti:2008cx}: Let $(\mathbb{R}^4,\eta)$ and $(M, \overline{g})$ be  respectively Minkowski spacetime and a FRW spacetime such that, wherever $M$ and $\mathbb{R}^4$ are diffeomorphic, $\overline{g}=a^2 \eta$, where $a$ is here an arbitrary smooth scale factor. Let $\frac{U(x,y)}{\sigma_M(x,y)}$ be the dominant divergence of the Hadamard parametrix in $M$. We omit the explicit $\epsilon$-dependence since it plays no role at this stage and it would only make the notation heavier. Then it holds that 
$$\frac{U(x,x^\prime)}{\sigma_M(x,x^\prime)}=\frac{1}{a(x)a(x^\prime)\sigma_{\mathbb{R}^4}(x,x^\prime)}+s(x,x^\prime),$$
where $s(x,x^\prime)$ is a less singular contribution to the Hadamard parametrix, while $\sigma_{\mathbb{R}^4}$ is the half squared geodesic distance in Minkowski spacetime. In other words, in order to evaluate the leading contribution to \eqref{Powerspectrum} we can use the following approximation for the two-point function of a Hadamard state on a Friedmann-Robertson-Walker spacetime with flat spatial sections and a scale factor constrained by \eqref{conf_fac}:
\begin{equation}\label{approx_2pt}
\omega_2(x,x^\prime)= \lim\limits_{\epsilon\to 0}\frac{1}{8\pi^2}\frac{a^{-1}(\tau)a^{-1}(\tau^\prime)}{\sigma_{\mathbb{R}^4}(x,x^\prime)+2i\epsilon(\tau-\tau^\prime)+\epsilon^2}+\textrm{less singular terms},
\end{equation}
where we have implicitly used the standard coordinates on $(M,g)$, namely $x=(\tau,\vec{x})$, $x^\prime=(\tau^\prime,\vec{x}^\prime)$ and $T(x)=\tau$, $T(x^\prime)=\tau^\prime$. Notice that this expression coincides with that employed in \cite{Pinamonti:2013zba} when $(M,g)$ is the cosmological de Sitter spacetime and, hence, $a(\tau)\propto\tau^{-1}$. For notational simplicity we shall write henceforth
\begin{equation}\label{approx_2pt_Mink}
\omega^2_{\mathbb{R}^4}(x,x^\prime)\doteq\lim\limits_{\epsilon\to 0}\frac{1}{8\pi^2}\frac{1}{\sigma_{\mathbb{R}^4}(x,x^\prime)+2i\epsilon(\tau-\tau^\prime)+\epsilon^2}.
\end{equation}
Notice that, by taking into account the most singular contribution to the two-point function, we are effectively focusing our attention on the ultraviolet behaviour of the state. In other words this amounts heuristically to considering large values of $a(\tau) k$. Although the power spectrum of phenomena such as the cosmic microwave background focus on the low momenta behaviour, we remark that, within the regime of validity of our approximation, we are not too far off the goal since the presence of the scale factor entails that we are discarding values of $k$ which are usually so low, not to be accessible for all practical purposes. Furthermore we shall show that, in the limit as $\tau\to -\infty$, the approximation tends to become exact and, thus, we reckon that our approximation can be safely used for early cosmological times.

\subsection{First order corrections to the power spectrum}

We have all ingredients to compute the power spectrum \eqref{Powerspectrum} taking into account the corrections due to the metric not being exactly the de Sitter one. As stressed already several times, we will be interested only in the leading singularity. In other words, on account of \eqref{approx_2pt}, we can switch from $\mathcal{P}(\tau,k)$ to 
\begin{equation}\label{P0}
\mathcal{P}_0(\tau,k)=m^4\int\limits_{-\infty}^\tau\int\limits_{-\infty}^\tau d\tau_1 d\tau_2\;\widehat\Delta_R(\tau,\tau_1,\vec{k})\widehat\Delta_R(\tau,\tau_2,\vec{k})a^{-2}(\tau_1)a^{-2}(\tau_2)\widehat{\omega}^2_{\mathbb{R}^4}(\tau_1,\tau_2,\vec{k}),
\end{equation}
where $\widehat{\omega}^2_{\mathbb{R}^4}(\tau_1,\tau_2,\vec{k})$ is the Fourier transform of \eqref{approx_2pt_Mink} along spatial directions. In order to compute $\mathcal{P}_0(\tau,k)$, we can use \eqref{ret_Green} to write explicitly the retarded Green operator as a Dyson series. We shall focus our attention on the first order corrections which can be evaluated using \eqref{Green_1st}. Inserting this expression in \eqref{P0} we get:
\begin{subequations}
\begin{flalign}
 & \mathcal{P}_0(\tau,k)  =  \mathcal{P}^{(0)}_0(\tau,k)+2\mathcal{P}^{(1)}_0(\tau,k) + \textrm{higher order corrections}\label{sum}\\
 & \mathcal{P}^{(i)}_0(\tau,k) =m^4\int\limits_{-\infty}^\tau\int\limits_{-\infty}^\tau d\tau_1 d\tau_2\;D_i(\tau,\tau_1,\tau_2,k) a^{-2}(\tau_1)a^{-2}(\tau_2)\widehat{\omega}^2_{\mathbb{R}^4}(\tau_1,\tau_2,\vec{k}),\;\;i=0,1\label{singleterms}
\end{flalign}
\end{subequations}
where 
\begin{subequations}
\begin{flalign}
 & D_0(\tau,\tau_1,\tau_2,k) \doteq \frac{a^4(\tau_1)a^4(\tau_2)}{36 a^4(\tau)}\widehat{\Delta}^0_R(\tau,\tau_1,k)\widehat{\Delta}^0_R(\tau,\tau_2,k) \label{D0}\\
& D_1(\tau,\tau_1,\tau_2,k) \doteq -\frac{a^4(\tau_1)a^4(\tau_2)}{36 a^4(\tau)}\widehat{\Delta}^0_R(\tau,\tau_1,k)\int\limits_{-\infty}^\tau dt\;\widehat{\Delta}^0_R(\tau,t,k)V(t)\widehat{\Delta}^0_R(t,\tau_2,k) ,\label{D1}
\end{flalign}
\end{subequations}
where $\widehat{\Delta}^0_R$ is the retarded Green function for the unperturbed operator $\frac{\partial^2}{\partial\tau^2}+\frac{k^2}{3}$. The next step in our investigation is the identification of a suitable bound for both terms appearing in \eqref{sum}. We shall analyze them separately.

\vskip.2cm

\noindent{\it A bound for $\mathcal{P}^{(0)}_0(\tau,k)$}: This is the leading term contributing to the singular structure of the power spectrum $\mathcal{P}(\tau,k)$ and, therefore, we must derive an estimate which is consistent with that of \cite{Pinamonti:2013zba} in the limit when $a(\tau)\propto\tau^{-1}$. For this reason we shall also adapt the computing method of Pinamonti and Siemssen to the case at hand. We remark once more that, in the case of a perfectly de Sitter background $\mathcal{P}^{(0)}_0(\tau,k)$ is the only contributing factor, contrary to the case we consider where additional terms appear. Inserting both \eqref{D0} and \eqref{unp_Green} in \eqref{singleterms}, it yields:
\begin{equation}\label{P0b}
\mathcal{P}^{(0)}_0(\tau,\kappa)=\lim\limits_{\epsilon\to 0}\frac{m^4}{288\pi^2}\int\limits_{\sqrt{3} \kappa}^\infty dp\;\frac{1}{\kappa^4}|A(\tau,\kappa,p)|^2e^{-\epsilon p},\quad\kappa\doteq \frac{k}{\sqrt{3}},
\end{equation}
where 
\begin{equation}\label{A} 
A(\tau,\kappa,p)\doteq\int\limits_{-\infty}^\tau d\tau_1\;\frac{\kappa a^2(\tau_1)}{a^2(\tau)}\sin\left(\kappa(\tau-\tau_1)\right)e^{-ip\tau_1}.
\end{equation}
As a first step we insert in this last expression the identity $e^{-ip\tau_1}=\left(\frac{d^2}{d\tau_1^2}+\kappa^2\right)\frac{e^{-ip\tau_1}}{\kappa^2-p^2}$, so that, after two integrations by parts, one obtains
$$A(\tau,\kappa,p)=\frac{\kappa^2}{\kappa^2-p^2}\left(e^{-ip\tau}+R(\tau,\kappa,p)\right),$$
where, omitting the dependence on $\tau_1$ of the scale factor,  
$$R(\tau,\kappa,p)=\frac{1}{a^2(\tau)}\int\limits_{-\infty}^\tau d\tau_1\;\left(\frac{2}{\kappa}(\ddot{a}a+\dot{a}^2)\sin(\kappa(\tau-\tau_1))-4\dot{a}a\cos(\kappa(\tau-\tau_1))\right)e^{-ip\tau_1}.$$
By using the bounds, valid for all $x\in\mathbb{R}$, $|\sin(x)|\leq |x|$ and $|\cos(x)|\leq 1$, as well as the hypothesis \eqref{conf_fac}, we get the following estimate uniform in $\tau$:
$$|R(\tau,\kappa,p)|\leq \frac{6}{a^2(\tau)}\int\limits_{-\infty}^\tau d\tau_1\;\dot{a}(\tau_1) a(\tau_1)=3.$$ 
In other words we have proven that, {\it uniformly in $\tau$}, 
\begin{equation}\label{A-bound}
|A(\tau,\kappa,p)|\leq \frac{4\kappa^2}{\left|\kappa^2-p^2\right|}.
\end{equation}
In order to get a result consistent with that of \cite{Pinamonti:2013zba}, we need also the behaviour of $|A(\tau,\kappa,p)|$ as $\tau$ diverges. In particular it holds that 
$$\lim\limits_{\tau\to -\infty}|A(\tau,\kappa,p)|=\frac{\kappa^2}{\left|\kappa^2-p^2\right|},$$
which is tantamount to proving that $|R(\tau,\kappa,p)|$ vanishes as $\tau\to -\infty$. This statement can be proven introducing the variable $x\doteq \frac{\tau}{\tau_1}$ in the integrand of $R(\tau,\kappa,p)$ and evaluating:
$$|R(\tau,\kappa,p)|\leq\frac{1}{a^2(\tau)}\left|\int\limits_{1}^\infty dx\;\left[4a^\prime a\cos(k\tau(1-x))-\frac{2}{\kappa\tau}(a^{\prime\prime}a+(a^\prime)^2)\sin(\kappa\tau(1-x))\right]e^{ip\tau x}\right|,$$
where the explicit dependence of $a$ form $x$ has been omitted and where the prime symbol indicates a derivation in $x$. While the contribution due to the second term in the integrand vanishes as $\tau\to -\infty$ on account of the decay properties of the scale factor, {\it c.f.} \eqref{conf_fac}, that of the first term tends to $0$ on account of the Riemann-Lebesgue lemma since the integrand, including the factor $a^{-2}(\tau)$ lies in $L^1[1,\infty)$. 

We have all the ingredients needed to evaluate \eqref{P0b}, namely, inserting \eqref{A-bound}, we can follow the same procedure as in \cite[Prop. 4.2]{Pinamonti:2013zba} to conclude
\begin{equation}\label{1st_estimate}
\left|\mathcal{P}^{(0)}_0(\tau,k)\right|\leq\frac{m^4}{2\pi^2}\int\limits_{k}^\infty dp\;\frac{1}{\left(3p^2-k^2\right)^2}=\frac{3-2\sqrt{3}\coth^{-1}(\sqrt{3})}{24\pi^2}\frac{m^4}{k^3},
\end{equation}
where we have restored the variable $k$. Furthermore, if we recall the asymptotic behaviour of $|A(\tau,\kappa,p)|$, this last expression entails that
$$\lim\limits_{\tau\to -\infty}\mathcal{P}^{(0)}_0(\tau,k)=\frac{m^4}{32\pi^2}\int_k^{+\infty}\left(\frac{1}{3p^2-k^2}\right)^2dp=\frac{3-2\sqrt{3}\coth^{-1}(\sqrt{3})}{384\pi^2}\frac{m^4}{k^3}.$$
To summarize we have proven the following:
\proposizione\label{prop1}{The leading contribution $\mathcal{P}^{(0)}_0(\tau,k)$ to the power spectrum of $\Psi$ induced by a conformally coupled, massive scalar field on an asymptotically de Sitter spacetime in the sense of \eqref{conf_fac} is bounded, uniformly in the conformal time, by a spectrum consistent with that of Harrison-Zel'dovich,
$$\left|\mathcal{P}^{(0)}_0(\tau,k)\right|\leq \frac{16 C}{k^3}\quad C=\frac{3-2\sqrt{3}\coth^{-1}(\sqrt{3})}{384\pi^2}m^4,$$
and it tends to it in the asymptotic limit, that is
$$\lim\limits_{\tau\to -\infty}\mathcal{P}^{(0)}_0(\tau,k)=\frac{C}{k^3}.$$}
Notice, moreover, that, in the limit where $a(\tau)\propto\tau^{-1}$ we obtain consistently the same result obtained by Pinamonti and Siemssen.

\vskip.2cm

\noindent{\it A bound for $\mathcal{P}^{(1)}_0(\tau,k)$}: Our goal is to find an estimate for the first order correction to the power spectrum. Inserting both \eqref{D1} and \eqref{unp_Green} in \eqref{singleterms}, it yields:
\begin{equation}\label{P1b}
\mathcal{P}^{(1)}_0(\tau,\kappa)=-\lim\limits_{\epsilon\to 0}\frac{m^4}{288\pi^2}\int\limits_{\sqrt{3}\kappa}^\infty dp\;\frac{1}{\kappa^4}A(\tau,\kappa,p)B(\tau,\kappa,p)e^{-\epsilon p},
\end{equation}
where we have used once more $\sqrt{3}\kappa=k$. The function $A(\tau,\kappa,p)$ is the same as in \eqref{A} whereas
\begin{subequations}
\begin{flalign}\label{B}
B(\tau,\kappa,p)=\int\limits_{-\infty}^\tau dt\; V(t)\sin\left(\kappa(\tau-t)\right)I(\tau,t,\kappa,p),\\
I(\tau,t,\kappa,p)=\int\limits_{-\infty}^\tau d\tau_1\; \frac{a^2(\tau_1)}{a^2(\tau)}\sin\left(\kappa(t-\tau_1)\right)e^{ip\tau_1}.\label{I}
\end{flalign}
\end{subequations}
In order to understand the behaviour of $B(\tau,\kappa,p)$ we start giving a closer look at $I(\tau,t,\kappa,p)$. By using once more the identity $e^{ip\tau_1}=\left(\frac{d^2}{d\tau_1^2}+\kappa^2\right)\frac{e^{ip\tau_1}}{\kappa^2-p^2}$ and by integrating twice by parts, we obtain that
$$|I(\tau,t,\kappa,p)|\leq\frac{\kappa^2}{|\kappa^2-p^2|}\left(\frac{|p|}{\kappa^2}+\frac{2\dot{a}(\tau)}{\kappa^2 a(\tau)}+\frac{1}{\kappa}+S(\tau,t,\kappa,p)\right),$$
where 
$$S(\tau,t,\kappa,p)=\frac{1}{a^2(\tau)}\int\limits_{-\infty}^t d\tau_1\;\left[\frac{2}{\kappa^2}\left(\ddot{a}a+\dot{a}^2\right)\sin\left(\kappa(t-\tau_1)\right)-\frac{4}{\kappa}\dot{a}a\cos\left(\kappa(t-\tau_1)\right)\right]e^{ip\tau_1}.$$
Notice that the explicit dependence of the scale factor on $\tau_1$ has been omitted in the integrand. Using the property that the absolute value both of the sine and of the cosine is bounded by $1$, it holds after integration that 
$$|S(\tau,t,\kappa,p)|\leq\frac{2}{\kappa}+\frac{2}{\kappa^2}\frac{\dot{a}(\tau)}{a(\tau)},$$
from which it descends
$$|I(\tau,t,\kappa,p)|\leq \frac{\kappa^2}{|\kappa^2-p^2|}\left(\frac{|p|}{\kappa^2}+\frac{4\dot{a}(\tau)}{\kappa^2 a(\tau)}+\frac{3}{\kappa}\right).$$
By inserting this estimate in \eqref{P1b}, we obtain eventually the following bound:
\begin{equation}\label{B-bound}
\left|B(\tau,\kappa,p)\right|\leq \frac{\kappa^2}{|\kappa^2-p^2|}\left(\frac{|p|}{\kappa^2}+\frac{4\dot{a}(\tau)}{\kappa^2 a(\tau)}+\frac{3}{\kappa}\right)\widetilde V(\tau),\qquad\widetilde V(\tau)\doteq\int\limits_{-\infty}^\tau dt\; |V(t)|,
\end{equation}
which, on account of \eqref{conf_fac} and \eqref{potential} entails also the $B(\tau,\kappa,p)\to 0$ as $\tau\to -\infty$. This is a consistency check with respect to the case of an exactly de Sitter spacetime. In this case $V(t)=0$ and therefore $\mathcal{P}^{(1)}_0(\tau,\kappa)=0$. 

We can now put together all ingredients, \eqref{A-bound} and \eqref{B-bound} in particular, to derive the following estimate for \eqref{P1b}:
\begin{equation*}
\begin{split}
\left|\mathcal{P}^{(1)}_0(\tau,k)\right|&\leq\frac{3m^2}{8\pi^2}\int\limits_k^\infty dp\;\left(\frac{1}{3p^2-k^2}\right)^2\left(\frac{p}{k^2}+\frac{4\dot{a}(\tau)}{k^2a(\tau)}+\frac{\sqrt{3}}{k}\right)\widetilde V(\tau)\\
&\leq \frac{m^4}{32\pi^2}\widetilde V(\tau)\left[\frac{1}{k^4}+\left(3-2\sqrt{3}\coth^{-1}(\sqrt{3})\right)\left(\frac{4}{k^5}\frac{\dot{a}(\tau)}{a(\tau)}+\frac{\sqrt{3}}{k^4}\right)\right]
\end{split}
\end{equation*}
where we have restored the variable $k=\sqrt{3}\kappa$. In other words we have found the following estimate:
\begin{empheq}[left=\empheqlbrace]{equation}
  \begin{split}\label{c1}
&\left|\mathcal{P}^{(1)}_0(\tau,k)\right|\leq \frac{c_1(\tau)}{k^4}+\frac{c_2(\tau)}{k^5},\\
&c_1(\tau)= \frac{m^4}{32\pi^2}\left(3\sqrt{3}+1-6\coth^{-1}(\sqrt{3})\right)\widetilde{V}(\tau),\\
&c_2(\tau)=\frac{m^4}{8\pi^2}\left(3-2\sqrt{3}\coth^{-1}(\sqrt{3})\right)\frac{\dot{a}(\tau)}{a(\tau)}\widetilde{V}(\tau).
\end{split}
\end{empheq}
Notice that, in view of the estimate for the leading order, {\it c.f.} Proposition \ref{prop1}, the bound that we have obtained is consistent with a sub-leading term only for large values of $k$ which are nonetheless the most interesting ones from a physical point of view. Yet we can show that there exists another estimate which shows that even for small values of $k$ no problem occurs. As a matter of facts, if we start once again from \eqref{B} and \eqref{I} and if we bound the absolute value of all trigonometric functions by $1$, we obtain that 
\begin{equation}\label{c3}
\left|B(\tau,k,p)\right|\leq c_3(\tau)\doteq\frac{\widetilde{V}(\tau)}{a^2(\tau)}\int\limits_{-\infty}^\tau d\tau_1\; a^2(\tau_1).
\end{equation}
This inequality, together with \eqref{A-bound}, both inserted in \eqref{P1b} yield
\begin{equation}\label{bound2}
\left|\mathcal{P}^{(1)}_0(\tau,k)\right|\leq\frac{m^4}{32\pi^2}\frac{c_3(\tau)}{k^2}\int\limits_k^\infty dp\;\frac{1}{3p^2-k^2}=\frac{m^4}{32\pi^2}c_3(\tau)\frac{\coth^{-1}(\sqrt{3})}{\sqrt{3}}\frac{1}{k^3}.
\end{equation} 
To summarize we have proven that 
\proposizione\label{prop2}{The first order contribution to the power spectrum for $\Psi$ of a conformally coupled, massive, Klein-Gordon field is bounded for large values of $k$ by 
$$\left|\mathcal{P}^{(1)}_0(\tau,k)\right|\leq \frac{c_1(\tau)}{k^4}+\frac{c_2(\tau)}{k^5},$$
where $c_1(\tau)$ and $c_2(\tau)$ are give in \eqref{c1} whereas, for small values of $k$
$$\left|\mathcal{P}^{(1)}_0(\tau,k)\right|\leq \frac{m^4}{32\pi^2}c_3(\tau)\frac{\coth^{-1}(\sqrt{3})}{\sqrt{3}}\frac{1}{k^3},$$
where $c_3(\tau)$ is give in \eqref{c3}. This last estimate entails, moreover, that
$$\lim\limits_{\tau\to -\infty}\left|\mathcal{P}^{(1)}_0(\tau,k)\right|=0.$$}

\vskip.2cm

\noindent{\it The first order bound for $\mathcal{P}_0(\tau,k)$}: We have now all ingredients to give a bound on the leading singular behaviour of the power spectrum $\mathcal{P}(\tau,k)$ under the assumption that we consider the effect of the departure from a perfectly de Sitter Universe as yielding a perturbation potential \eqref{potential}, whose effect we consider at first order in a perturbative series. More precisely, switching from $\mathcal{P}(\tau,k)$ to \eqref{P0} and then, using \eqref{sum} discarding all higher order terms, we obtain
$$|\mathcal{P}_0(\tau,k)|\leq|\mathcal{P}^{(0)}_0(\tau,k)|+2|\mathcal{P}^{(1)}_0(\tau,k)|.$$
We can now use both Proposition \ref{prop1} and Proposition \ref{prop2} to conclude the following:
\teorema\label{main}{At first order in the perturbative series for the potential $V(t)$ as in \eqref{potential}, the power spectrum of  $\Psi$ induced by a conformally coupled, massive scalar field is bounded for large values of $k$ by
$$|\mathcal{P}_0(\tau,k)|\leq\frac{16C}{k^3}+\frac{2c_1(\tau)}{k^4}+\frac{2c_2(\tau)}{k^5},$$
where $c_1(\tau)$ and $c_2(\tau)$ are given in \eqref{c1} while $C$ in Proposition \ref{prop1}. On the contrary when $k$ is small and, hence, its higher inverse powers are dominant, the following bound holds:
$$|\mathcal{P}_0(\tau,k)|\leq\frac{\widetilde{C}(\tau)}{k^3},$$
where $\widetilde{C}(\tau)\doteq 16C+2c_3(\tau)$ where $c_3(\tau)$ is given in \eqref{c3}. In both cases
$$\lim\limits_{\tau\to -\infty}|\mathcal{P}_0(\tau,k)|=\frac{C}{k^3}.$$
}
Let us comment on our final result: The hypothesis that the spacetime is not exactly de Sitter but tends to it asymptotically in the past appears to modify the computation of the power spectrum for the scalar fluctuations of the metric in a rather controlled form. As a matter of fact the two-point correlation function gets altered by the appearance of a retarded fundamental solution which is built out of a wave-like equation with a non trivial time-dependent potential. If we consider this one as a perturbation we have shown that, while at zeroth order we recover, as expected, the same result obtained in \cite{Pinamonti:2013zba}; the first order correction contributes with terms which are for large $k$ much smaller. Furthermore, the time dependence of the coefficients is under control insofar that the limit as the conformal time tends to $-\infty$ reproduces exactly the result in de Sitter spacetime. To strengthen the result one has to remark that, although, the coefficients for the terms $k^{-4}$ and $k^{-5}$ can be in principle quite big, nonetheless, we have also exhibit a bound, which is always valid (although we have associated it to small values of $k$) and which shows that the first order correction always behaves at worst consistently with the Harrison-Zel'dovich power spectrum. 

Of course the question on the effect of the higher order contributions has not been tackled. Let us stress nonetheless that these cases could be treated repeating the exact same steps as for $\mathcal{P}^{(1)}_0(\tau,k)$. As a matter of fact one would have to evaluate more integrals containing the perturbation potential, but, always on account of the fact that $V(\tau)\to 0$ as $\tau\to -\infty$, we would always find contributions due to smaller and smaller inverse powers of $k$. In this sense the conclusions of Theorem \ref{main} are of wider validity. More complicated is instead finding an answer to the question whether the ensuing perturbation series for the power spectrum is convergent. We will not tackle the problem here although preliminary analyses suggest that it is indeed the case.

\se{Conclusions}
The goal of our paper was to show the robustness of the method proposed in \cite{Pinamonti:2013zba} by extending its range of applicability to a wider class of spacetimes, which are less ideal than the flat patch of the de Sitter Universe and more closely related to the actual physical scenario. As a matter of fact we have shown that the effect of the scale factor \eqref{conf_fac} is to introduce a non-trivial potential in the relevant equations. In order to obtain an estimate for the power spectrum we had to treat such potential perturbativly, a fact which, in turn, yields the power spectrum itself as a Dyson series. We have shown that, despite we are dealing with a simple toy model, the leading order behaves consistently with the Harrison-Zel'dovich spectrum, while the first order is sub-dominant and it actually vanishes as $\tau\to-\infty$.
In other words, encoding at a geometric level that the spacetime behaves as a de Sitter Universe only asymptotically in time, entails
that perfect scale invariance of the power spectrum cannot hold true. All our results are thus not in contrast with the present experimental predictions and, as a by-product, they also show the efficiency of the algebraic approach to quantum field theory also to compute concrete quantities in the framework of cosmology and not not only to unveil rigorous, structural properties -- in this context, see also \cite{Eltzner:2013soa, Fredenhagen:2013vxa, Hack:2013aya}.

Of course several problems are left open. From a more formal point of view, the method we have used allows us to control each term in \eqref{sum}, although we have only written explicitly the first order. What is lacking is an answer on whether the whole series converges. Preliminary investigations suggest that this is indeed the case, but such issue deserves a thorough analysis. From a more physical point of view we have considered only scalar fluctuations of the metric. Hence it would be desirable to try to extend the whole approach also for the tensor fluctuations, although one has to keep in mind that the difficulty of this investigation is much higher than the present one, on account in particular of gauge invariance. Yet we feel that this is the natural following step and we hope to be able to report on it in the next future.

\section*{Acknowledgements}
We would like to thank Nicola Pinamonti and Daniel Siemssen for enlightening discussions, particularly on \cite{Pinamonti:2013zba}. We are grateful to Daniel Siemssen for useful comments on the final version of this manuscript. This paper is based on the master thesis of A.M. {\it ``Curvature Fluctuations in Asymptotically de Sitter Spacetimes''}, defended at the University of Pavia on the 30/04/2014. The work of C.D. has been supported partly by the University of Pavia  and partly by the Indam-GNFM project {\it ``Influenza della materia quantistica sulle fluttuazioni gravitazionali''}. C. D. gratefully acknowledges the kind hospitality of the Erwin Schr\"odinger Institute during the workshop {\it ``Algebraic Quantum Field Theory: Its Status and Its Future''}.

\end{document}